\documentclass[]{main}

\usepackage{tikz}

\usetikzlibrary{
    arrows.meta,
    positioning,
    shapes.geometric,
    shapes.misc,
    fit,
    calc,
    intersections,
    decorations.pathreplacing,
    arrows.meta
}

\tikzset{
    >={Latex},
    block/.style={draw, rounded corners=3pt, fill=white, align=center, minimum width=0.38\textwidth, minimum height=9mm},
    smallblock/.style={draw, rounded corners=2pt, fill=white, align=center, minimum width=0.33\textwidth, minimum height=8mm},
    decision/.style={draw, diamond, aspect=2.2, inner sep=1.2pt, align=center},
    note/.style={draw, fill=gray!10, rounded corners=2pt, font=\small, inner sep=2pt},
    parbox/.style={draw, dashed, rounded corners=3pt, inner sep=4pt},
    lbl/.style={font=\footnotesize\itshape, inner sep=1pt, anchor=west},
    line/.style={-Latex, thick}
}

\tikzset{every node/.style={font=\small}}

\definecolor{grey}{RGB}{90,90,90}

\begin{document}

\IACpaperyear{2025}

\IACpapernumber{IAC-25,B2,7,9,x99345}

\IAClocation{Sydney, Australia}

\IACdate{29 Sep-3 Oct 2025}

\IACcopyrightB{Andrew Savchenko.}

\title{Authenticated encryption for space telemetry}

\IACauthor{Andrew Savchenko$^{\orcidlink{0009-0009-4030-3667}}$}{1}{1}

\IACauthoraffiliation{University of Technology Sydney, Data Science Institute,
Behavioural Data Science Lab. 61 Broadway, Ultimo, NSW 2007.
\normalfont{E-mail:~\authormail{andrew.savchenko@student.uts.edu.au}}
}

\abstract{
    \hspace{11pt}We explore how command stack protection requirements outlined
    in NASA-STD-1006A can be satisfied within the context of emergency space
    telemetry. The proposed implementation of lightweight authenticated encryption
    offers strong security without sacrificing performance in
    resource-constrained environments. It produces fixed-length messages,
    maintaining compatibility with the underlying data transport protocols. By
    focusing on predictable properties and robust authentication, we create a
    scheme that protects the confidentiality, integrity and authenticity of
    telemetry data in emergency communications while balancing security
    requirements with the operational constraints.\\

}

\maketitle

\thispagestyle{fancy}

\section*{Acronyms/Abbreviations}

\begin{description}
        [
            font=\normalfont,
            style=standard,
            itemsep=0pt,parsep=1pt,
            labelsep=0em,
            leftmargin=1.7cm,
            labelwidth=1.7cm,
        ]
        \begin{raggedright}
        \item[AAD]   Additional Authenticated Data
        \item[AES]   Advanced Encryption Standard
        \item[BCH]   Bose Chaudhuri Hocquenghem
        \item[CCSDS] Consultative Committee, Space Data Systems
        \item[CRC]   Cyclic Redundancy Check
        \item[CRL]   Certificate Revocation List
        \item[DEK]   Data Encryption Keys
        \item[DTN]   Disruption Tolerant Networking
        \item[ECDHE] Diffie-Hellman Ephemeral
        \item[FEC]   Forward Error Correction
        \item[FIPS]  Federal Information Processing Standards
        \item[GCM]   Galois Counter Mode
        \item[GHASH] Galois Hash
        \item[GNSS]  Global Navigation Satellite System
        \item[IV]    Initialisation Vector
        \item[KEK]   Key Encryption Keys
        \item[NTP]   Network Time Protocol
        \item[OCSP]  Online Certificate Status Protocol
        \item[PPM]   Parts Per Million
        \item[RTC]   Real Time Clock
        \item[SASIC] South Australian Space Industry Centre
        \item[SGB]   Second Generation Beacon
        \item[SoC]   System On Chip
        \item[SPP]   Space Packet Protocol
        \item[SW]    SpaceWire
        \end{raggedright}
\end{description}

\section{Introduction}

\subsection{Objectives}

Our objective was to create an encryption scheme for fixed-size Second
Generation Beacon (SGB) messages that is independent from the underlying
transport and compatible with Space Packet Protocol (SPP), SpaceWire (SW) and
Delay / Disruption Tolerant Networking (DTN)\footnote{Format compatibility is
preserved, however the 2 sec. acceptance window discussed later must be relaxed
for DTN store-and-forward transmissions.}.

NASA-STD-1006A requirement \cite{nasa-std-1006a-2022} to protect command stack
with FIPS-140 level 1 encryption along with the strong recommendation
\cite{ccsds-352-0-b-2-2019} to use authenticated encryption issued by
Consultative Committee for Space Data Systems (CCSDS), motivated us to design a
simple and durable routine with predictable properties that is easy to
understand and implement across various platforms.

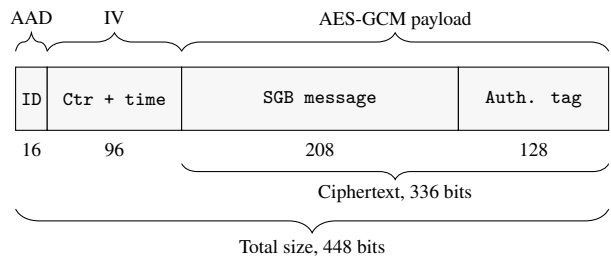
\begin{figure}[H]
    \newcommand{\figscale}{1.051}
    \begin{tikzpicture}[
            box/.style={draw, minimum height=0.8cm, inner sep=0pt, outer sep=0pt, font=\scriptsize},
            label/.style={font=\scriptsize},
            dimension/.style={<->, >=stealth, thick},
            scale=\figscale, transform shape
        ]
        \def\aadwidth{0.4}
        \def\ivwidth{1.7}
        \def\payloadwidth{3.5}
        \def\tagwidth{1.9}
        \def\boxheight{0.8}
        \node[
            box,
            minimum width=\aadwidth cm,
            minimum height=\boxheight cm,
            fill=grey!5
        ]
        (aad) at (0,0) {\scriptsize \texttt{ID}};
        \node[
            box,
            minimum width=\ivwidth cm,
            minimum height=\boxheight cm,
            fill=grey!5,
            right=0pt of aad
        ]
        (iv) {\scriptsize \texttt{Ctr + time}};
        \node[
            box,
            minimum width=\payloadwidth cm,
            minimum height=\boxheight cm,
            fill=grey!5,
            right=0pt of iv
        ]
        (payload) {\scriptsize \texttt{SGB message}};
        \node[
            box,
            minimum width=\tagwidth cm,
            minimum height=\boxheight cm,
            fill=grey!5,
            right=0pt of payload
        ]
        (tag) {\scriptsize \texttt{Auth. tag}};
        \node[label, below=1pt of aad] {16};
        \node[label, below=1pt of iv] {96};
        \node[label, below=1pt of payload] {208};
        \node[label, below=1pt of tag] {128};
        \draw[
            decorate,
            decoration={brace, amplitude=7pt, raise=6pt}
        ]
        (aad.north west) -- (aad.north east)
        node[midway, above=12pt, label] {AAD};
        \draw[
            decorate,
            decoration={brace, amplitude=7pt, raise=6pt}
        ]
        (iv.north west) -- (iv.north east)
        node[midway, above=12pt, label] {IV};
        \draw[
            decorate,
            decoration={brace, amplitude=7pt, raise=6pt}
        ]
        (payload.north west) -- (tag.north east)
        node[midway, above=10pt, label] {AES-GCM payload};
        \draw[
            decorate,
            decoration={brace, amplitude=9pt, raise=-7pt, mirror}
        ]
        ([yshift=-35pt]aad.south west) --
        node[below=1pt, label] {Total size, 448 bits} ([yshift=-35pt]tag.south east);
        \draw[
            decorate,
            decoration={brace, amplitude=6pt, raise=0pt, mirror}
        ]
        ([yshift=-12pt]payload.south west) -- ([yshift=-12pt]tag.south east)
        node[midway, below=4pt, label] {Ciphertext, 336 bits};
    \end{tikzpicture}
    \caption{Message structure showing the encryption format.}
    \label{fig:message_structure}
\end{figure}

\subsection{Background}

On Earth, SGB messages are intentionally transmitted unencrypted. However, in
space this would be detrimental for several reasons:

\begin{enumerate}[itemsep=0pt]
    \item If the message is not encrypted -- loss of confidentiality via disclosure
        of the sensitive telemetry. It might be not exploitable by itself, but
        inform an attacker about the condition of the asset and allow them to
        adjust their deployable effects and tactics.
    \item If the message is not signed -- loss of availability via denial attacks
        where an arbitrary sender is able to confuse and/or overwhelm the receiver
        with counterfeit messages at a low cost.
    \item If the message is encrypted, but not authenticated -- malleable encryption
        would allow an attacker to perform blind bit flips or targeted
        modifications. If such changes are not detected, it would subsequently
        cause loss of integrity with a wide range of potential consequences.
\end{enumerate}

Separately, we also need to consider replay attacks -- it should be impossible
to retransmit previously captured authentic messages and have them accepted by
the receiver.

\subsection{Existing literature}

Existing literature and industry standards \cite{ccsds-352-0-b-2-2019,
nasa-std-1006a-2022} strongly suggest the use of authenticated encryption while
the second issue of information report by CCSDS on cryptographic algorithms
specifically recommends \cite{ccsds3509g2} to rely on Advanced Encryption
Standard (AES) operating in \mbox{Galois / Counter} mode (GCM) with the standard
96-bit initialisation vector (IV).

\mbox{AES-GCM} is a well-tested algorithm used in MAC security, SSH, TLS, IPsec
and others. There are commercially available, system-on-chip (SoC) modules that
feature hardware AES accelerators, which is important in environments with
limited power supply and compute resources. Using hardware instruction sets for
implementing AES rounds does not only make the computation dramatically more
power-efficient, but also renders timing attacks highly impractical, especially
in environments where physical access to the host is limited
\cite{MoweryKeelveedhiShacham2012}.

There are, however, recent proposals that looked into more complex cryptographic
approaches. For example, some explore \cite{saleem2024-cospas-sarsat-ecdsa} use
of public key cryptography which offer certain theoretical advantages such as
non-repudiation and simplified key distribution. However, it introduces
operational complexity: an undesirable trait \cite{khandker2024-cospas-mac} in
circumstances when rapid updates of the software stack are not readily available
and additional computational complexity is unsuitable.

Others propose \cite{smailes2025keyspaceenhancingpublickey} to use public key
infrastructure (PKI) utilising terrestrial standards such as certificate
revocation lists (CRLs) and online certificate status protocol (OCSP). While
such approach offers rather sophisticated key management capabilities suitable
for complex environments, it introduces measurable (and potentially significant)
computational overhead and relies on certificate authorities that may be
inaccessible if infrastructure is under a jamming attack or simply degraded due
to force majeure.

In contrast, our symmetric key approach prioritises operational simplicity and
deterministic performance characteristics essential for the emergency telemetry.
It focuses specifically on authenticated encryption for fixed-format emergency
messages. Our scheme with pre-shared keys eliminates overheads such as
certificate validation, revocation checking, and periodic key rotation while
still meeting the NASA-STD-1006A requirements.

This fundamental difference in approach reflects different operational
priorities: some optimise for flexibility and scalability in heterogeneous
networks, while our solution aims to optimise for efficiency and reliability in
resource-constrained deployments where communicating parties are known in
advance and the message format is fixed. It aligns with the CCSDS
recommendations while maintaining implementation simplicity and providing both
confidentiality and integrity guarantees. The use of \mbox{AES-GCM} specifically
follows established precedent in secure communications protocols including TLS
1.3, IPSec, and SSH. The widespread adoption increases confidence in algorithm
security properties, offers a wide number of well-tested software libraries and
ensures broad hardware support.

Our implementation uses a combination of 64-bit timestamp and 32-bit counter to
create a 96-bit IV which is aligned with the CCSDS recommendations; we believe
that design choice is justified by the following considerations:

\begin{enumerate}[itemsep=0pt]
    \item Timestamps provide guaranteed uniqueness through the monotonic
        progression, thus eliminating the reuse risks. 64-bit space is
        sufficient -- even at 1 Hz messaging, exhaustion would require billions
        of years.
    \item The counter is incremented atomically on system restart and before any
        emergency transmission to mitigate the risk of primary real-time clock
        (RTC) reset. It can also act as the secondary RTC, continuously
        incrementing each second starting at T-Zero.
    \item AES-GCM requires IVs that are \(\neq96\) bit to be processed through
        the Galois hash (GHASH) function to derive the 128-bit initial counter.
        Our solution avoids this route and takes maximum advantage of the
        hardware paths.
\end{enumerate}

This approach maintains standards compliance while providing operational
efficiency and maintaining cryptographic security within the mission boundaries.

\section{System constraints and design considerations}

\subsection{Key limitations and assumptions}

We begin with acknowledging inherent limitations of space systems and then
derive specific operational constraints based on these findings.

\begin{enumerate}[itemsep=0pt]
    \item Power supply -- transmitter power is limited and declining over the
        mission lifetime due to solar array degradation, battery ageing, etc.
    \item Link -- constrained budget, in a LEO mission we account for the 9.6
        kb/s bandwidth via UHF downlink with link-layer forward error correction
        (FEC).
    \item Processing resources -- space-qualified or higher-grade automotive
        SoCs have limited computational capacity. Cryptographic operations must
        minimise CPU cycles to avoid impacting other critical systems.
    \item On-board memory -- sufficient to persistently store the counter value,
        ferroelectric or similar non-volatile.
    \item Ground contact window -- data transmission opportunities are limited
        to specific orbital passes, typically 5-15 minutes per contact which
        might be further reduced depending on the transmitter and receiver
        antennas, weather, and so forth.
\end{enumerate}

Given these constraints, we can derive specific operational parameters:

\begin{enumerate}[itemsep=0pt]
    \item Desirable payload size of $\approx450$ bits, to keep it under $5\%$ of
        the 9.6 kb/s budget, allowing over $95\%$ of the bandwidth to be
        allocated for other needs.
    \item Messaging rate $\leq1$ Hz. This ensures that we have adequate
        processing time for cryptographic operations while maintaining the
        overall system responsiveness.
\end{enumerate}

\subsubsection{Timing}\label{sec:timing}

Given the nature of space-qualified hardware, it is reasonable to assume that
both transmitter and receiver have high-precision, real-time clocks accurate to
hundreds of milliseconds or better. While receivers can rely on continuous
synchronisation over global navigation satellite system (GNSS), network time
protocol (NTP) and other means, a transmitter can utilise one of the many
\mbox{$\pm 3$} parts per million (PPM) commercial RTCs qualified for wide
temperature ranges.

For a LEO satellite, it is reasonable to assume that clock sync is possible at
least twice per day. We can then calculate time drift as: $ PPM \div 10^6 \times
Time_{sec} \times 1000 $. Therefore, \mbox{$\pm 3$} PPM RTC would give us a
drift of: $$ 3 \div 10^6 \times 43200 \times 1000 \approx 130\ ms $$ Even if we
add a full second to account for variable latency due to the weather and other
conditions, an opportunity window of 2 seconds should accommodate the clock
drift and rounding errors due to integer quantisation.

\section{Proposed scheme}

\subsection{Protocol overview}

The proposed scheme leverages AES-GCM, a well-tested cipher accelerated in hardware
across multiple platforms, making it suitable for space applications with
limited power and computational resources. The protocol uses symmetric
authenticated encryption with a shared secret key and 96-bit IV composed of
32-bit counter and 64-bit timestamp to ensure confidentiality, integrity and
replay protection.

While our scheme detects corruption and authenticates messages, it does not
correct channel errors. This separation follows CCSDS recommendation
\cite{CCSDS351} for space data systems, where synchronisation and channel coding
operate below the security layer.

\subsection{Message structure}

The encrypted message shown in (Fig.\,\ref{fig:message_structure}) maintains a
fixed length of 448 bits which consists of Additional Authenticated Data (AAD),
IV and ciphertext; latter further separated into the payload and tag.

\begin{enumerate}[itemsep=0pt]
    \item AAD -- 16-bit asset ID: unencrypted, but authenticated. Allows rapid
        identification and tracking of the transmitters to counter replays on
        receivers. 2 bytes covers 65,536 unique values, more than enough given
        the current number ($\approx15000$) of active satellites.
    \item IV -- 96 bits, combination of the current time and auxiliary counter
        $R[32] \parallel T[64]$. Serves as the nonce for the encryption and
        assists in detecting replays on receivers.
    \item Ciphertext -- 336 bits, which contains:
        \begin{enumerate}[itemsep=0pt]
            \item Encrypted payload -- 208 bits (SGB message, padded to the byte
                boundary\footnote{SGB messages are 202 bits $\div$ 8 = 25.25 bytes,
                hence we round to 26 bytes.}).
            \item Authentication tag -- 128 bits. As the cost of successful
                forgeries in emergency messages is high, we chose a conservative
                upper boundary. Moreover, CCSDS 352.0-B-2 explicitly states
                \cite{ccsds-352-0-b-2-2019} that `MAC tag size shall be 128 bits'.
        \end{enumerate}
\end{enumerate}

All multibyte integers used in the message structure (AAD, counter, timestamp)
must be treated as big-endian.

\subsubsection{Overhead and its justification}\label{sec:overhead}

448-bit frame represents a $\approx79\%$ overhead compared to the standard 250
bit SGB frame: 202 bits payload + 48 bits Bose-Chaudhuri-Hocquenghem (BCH) error
correction.

However, we assert that such increase is justified as we meet the NASA-STD-1006A
requirements \cite{nasa-std-1006a-2022} and end up with the predictably-sized
payload that can be transferred in a comparatively short time. Let us prove it;
we will use the following constants:

\begin{fleqn}
    \[
        \begin{aligned}
            \text{Altitude}\ (A) &\approx 500\ \text{km} \\
            \text{Data size } (D) &= 448\ \text{bits} \\
            \text{Carrier frequency} (F) &= 400\ \text{MHz} = 4 \times 10^8\ \text{Hz} \\
            \text{Link bandwidth}\ (L) &= 9.6\ \text{kb/s} = 9.6 \times 10^3\ \text{bps} \\
            \text{Speed of light}\ (S) &= 300000\ \text{km/s} = 3 \times 10^8\ \text{m/s} \\
        \end{aligned}
    \]
\end{fleqn}

\noindent Then we should calculate the propagation delay. A generic formula is:
$$ Time = {Distance}\ \div\ {Signal\ Speed} $$

\noindent We represent the altitude of \(500\ km \) as \(5 \times 10^{5}\ m\),
then the time for a one-way transmission becomes: $$ (5 \times 10^{5}\ m) \div
(3 \times 10^{8}\ m/s) \approx 1.67\ ms $$

\noindent Or, \( 1.67 \times 2 = 3.34\ ms\) for a roundtrip. Now we need to
account for the encrypted transmission time: $$ 448\ bits \div 9.6 \times
10^{3}\ bps \approx 46.67\ ms $$

\noindent In a similar way we can derive transmission time for the original, 250
bit SGB message: $$ 250\ bits \div (9.6 \times 10^{3}\ bps) \approx 26\ ms $$
This, finally, allows us to calculate the total one-way latency for the
encrypted payload: $$ 1.67\ {ms} + 46.67\ ms = 48.34\ ms $$

\noindent and for the regular SGB message: $$ 1.67\ {ms} + 26\ ms = 27.67\ ms $$

\noindent Even considering the high angular velocity of the LEO satellite in
relation to the terrestrial receiver, we assert that $20.67$ ms difference is
practically negligible. Let us prove this point.

To begin with, we need to find the distance that satellite would travel in the
given time. For this, we need to know time $(t)$ and orbital speed $(v)$. We
know that the satellite is on LEO orbit and thus can assume
\cite{esa_types_of_orbits} its speed to be \mbox{$\approx$ 7.8 km/s} relative to
Earth and orbit duration $(T)$ of 90 minutes. Then, \( (v) = 7.8 \times 10^3\
m/s\) and distance travelled becomes: $$ (v) \times (t) = (7.8 \times 10^3)\
m/s\ \times\ 0.02067\ s \approx 161.2\ m $$

\noindent We will take the worst-case scenario where the satellite is deployed
on retrograde orbit and account for the Earth rotation. We know that Earth
surface speed at the equator is $465$ m/s, consequently: $$ 465\ m/s \times 0.02067\
s \approx 9.6\ m $$

\noindent which gives us total offset of: $$ 161.2 + 9.6 = 170.8\ m $$

\noindent Now, we need to account for two separate issues:

\begin{enumerate}[itemsep=0pt]
    \item Doppler shift -- important, as the shift can push the signal outside
        of the narrowband range which can, at extremes, cause loss of lock and
        termination of the connection or just increase the error rate and cause
        loss of packets that will lead to retransmissions.
    \item Antenna coverage -- significant, as this is something we need to
        account for when using narrowly focused antennas on transmitter and
        receiver.
\end{enumerate}

\noindent Let us take care of the Doppler offset first. We know that the Doppler
shift formula is: $\Delta (F) = (v) \div (S) \times (F)$. Accounting for the
Earth speed, $(v)$ becomes \( 7800\ m/s + 465\ m/s = 8265\ m/s \) which allows
us to calculate the maximum Doppler shift as: $$ \Delta (F) = 8265 \div (3
\times 10^8) \times (4 \times 10^8) = 11020\ Hz $$

\noindent During a pass, the satellite transitions from positive to zero and
negative shift resulting in a total Doppler change of $2 \times 11020 = 22040$
Hz. We assume a 600 seconds pass as: $$ \text{Doppler rate} = 22040 \div 600
\approx 36.7\ Hz/s $$

\noindent which allows us to demonstrate an insignificant difference between the
regular SGB message and frame of the encrypted payload: $$ 36.7\ Hz/s \times
0.02067\ s \approx 0.76\ Hz $$

\noindent Now, let us address the 2\textsuperscript{nd} point concerning the
antenna coverage. We assume parabolic antenna with $10^{\circ}$ angle. We
compute beam footprint radius $(r)$ projected on the ground as: $$
(r) = altitude \times \tan(full\ beamwidth \div 2) $$

\noindent And with the actual numbers: $$ (r) = 500000 \times \tan(5^\circ) =
500000 \times 0.0875 = 43750\ m $$

\noindent or $\approx 43.8\ km$, which makes total offset of $\approx0.17$ km
look well and truly negligible in comparison.

In conclusion, we believe that these calculations clearly demonstrate how
inconsequential is the overhead despite the seemingly large relative
$\approx79\%$ increase in size.

\section{Cryptography}

In this section we will employ the following notations:

\begin{fleqn}
    \[
        \begin{aligned}
            &\textit{A} - \text{AAD for GCM, unique per asset} \\
            &\textit{C} - \text{Ciphertext} \\
            &\textit{K} - \text{Shared secret key, unique per asset} \\
            &\textit{P} - \text{Plaintext} \\
            &\textit{R} - \text{Monotonic counter, stored in non-volatile memory} \\
            &\textit{S} - \text{Replay state: } A \rightarrow (R_{last}, T_{last}^S) \\
            &\textit{T$^S$} - \text{Sender timestamp} \\
            &\textit{T$^R$} - \text{Receiver timestamp} \\
            &\underline{\text{E}} - \text{Encrypt} \\
            &\underline{\text{D}} - \text{Decrypt} \\
            &\underline{\text{V}} - \text{Verify}
        \end{aligned}
    \]
\end{fleqn}

The encryption process itself is rather straightforward:

\begin{enumerate}[itemsep=0pt]
    \item Generate timestamp and convert it to 64 bits binary on the
        transmitter, increment the 32-bit counter.
    \item Combine the counter with the timestamp into the IV.
    \item Initialise AES-GCM, encrypt payload.
    \item Concatenate AAD, IV and the resulting ciphertext.
\end{enumerate}

Which effectively becomes:

\begin{flalign}
    &IV = (R \parallel T^S) \\
    &A \parallel IV \parallel \text{AES-GCM}(K, IV, P, A)
\end{flalign}

Hence the decryption process is:

\begin{enumerate}[itemsep=0pt]
    \item Split the received message into:
        \begin{enumerate}
            \item AAD -- first 2 bytes.
            \item IV -- next 12 bytes.
            \item Ciphertext -- rest of the message.
        \end{enumerate}
    \item Extract counter (first 4 bytes) and timestamp (last 8 bytes) from the
        12-byte IV.
    \item Generate current timestamp $T^R$ on the receiver.
    \item Verify that the extracted timestamp $T^S$ is within the 2 second
        window.
    \item Assert the length of the ciphertext $|C|=42$ bytes.
    \item Check replay protection -- if AAD exists in state $S$, verify that $(R
        > R_{last}) \land (T^S > T_{last}^S)$.
    \item Decrypt using the $(K, IV, A)$.
    \item Update replay state $S$.
\end{enumerate}

This can be expressed in a concise manner as following:

\begin{flalign}
    &\{A, IV, C\} \text{ where } |A| = 2, |IV| = 12, |C| = 42 \\
    &R = IV_{0:4}\text{ and } T^S = IV_{4:12} \\
    &\text{If } A \in S\text{ then } (R_{last}, T_{last}^S) = S[A]  \\
    &\text{Reject if } A \in S \land (R \leq R_{last} \lor T^S \leq T_{last}^S) \\
    &\operatorname{\underline{V}}(T^S, T^R, C) = (|C| = 42) \land ((T^R - T^S) \leq 2) \\
    &\operatorname{\underline{D}}(K, IV, C, A) = \mathrm{AES\mbox{-}GCM}^{-1}(K, IV, C, A) \\
    &S[A] \leftarrow (R, T^S)
\end{flalign}

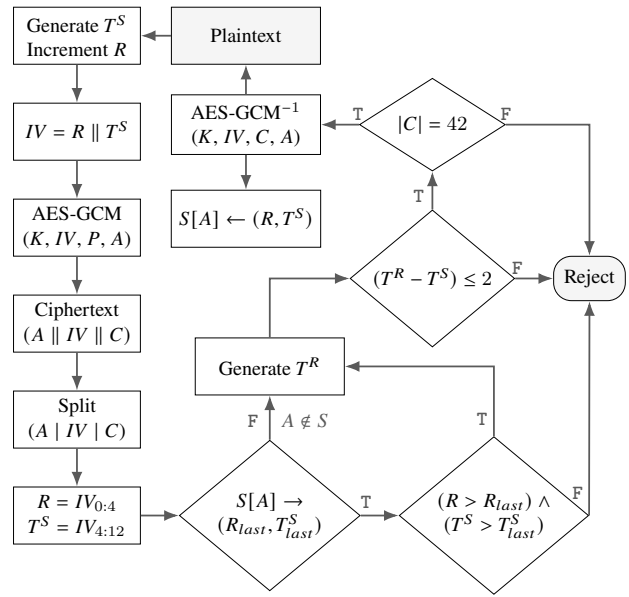
\begin{figure}[H]
    \resizebox{\columnwidth}{!}{%
        \begin{tikzpicture}[
                node distance=6mm,
                block/.style={
                    draw,
                    rounded corners=0pt,
                    fill=white,
                    align=center,
                    minimum width=20mm,
                    minimum height=9mm,
                    font=\small
                },
                data/.style={
                    draw,
                    fill=gray!8,
                    align=center,
                    minimum width=20mm,
                    minimum height=9mm,
                    font=\small
                },
                crypto/.style={
                    draw,
                    rounded corners=0pt,
                    fill=white,
                    align=center,
                    minimum width=20mm,
                    minimum height=9mm,
                    font=\small
                },
                decision/.style={
                    draw,
                    diamond,
                    aspect=1.2,
                    inner sep=2pt,
                    minimum width=23mm,
                    minimum height=10mm,
                    align=center,
                    font=\small
                },
                reject/.style={
                    draw,
                    rounded corners=8pt,
                    fill=gray!8,
                    align=center,
                    minimum width=11.3mm,
                    minimum height=7mm,
                    font=\small
                },
                state/.style={
                    draw,
                    dashed,
                    rounded corners=0pt,
                    fill=gray!5,
                    align=center,
                    minimum width=20mm,
                    minimum height=9mm,
                    font=\small
                },
                line/.style={->, thick, color=grey},
                return_arrow/.style={->, thick, dotted, color=grey},
            ]
            \begin{scope}[]
                \node[
                    data,
                    minimum width=23.5mm
                ] (payload1) {Plaintext};
                \node[
                    block,
                    left=4.75mm of payload1
                ] (timestamp1) {Generate $T^S$\\Increment $R$};
                \node[
                    block,
                    below=of timestamp1
                ] (counter) {$IV = R \parallel T^S$};
                \node[
                    crypto,
                    below=of counter
                ] (encrypt) {AES-GCM\\{($K$, $IV$, $P$, $A$)}};
                \node[
                    block,
                    below=of encrypt
                ] (concat) {Ciphertext\\$(A \parallel IV \parallel C)$};
                \node[
                    block,
                    below=of concat
                ] (split) {Split\\($A \mid IV \mid C$)};
                \node[
                    block,
                    below=of split
                ] (extract) {$R = IV_{0:4}$\\$T^S = IV_{4:12}$};
                \node[
                    decision,
                    right=of extract
                ] (checkstate) {$S[A] \rightarrow$\\$(R_{last}, T_{last}^S)$};
                \node[
                    decision,
                    right=of checkstate
                ] (replay) {$(R > R_{last})\ \land$\\ $(T^S > T_{last}^S)$};
                \node[
                    block,
                    minimum width=23.5mm,
                    above=7mm of checkstate
                ] (timestamp2) {Generate $T^R$};
                \node[
                    decision,
                    above=14mm of replay,
                    xshift=-9.7mm
                ] (check) {$(T^R - T^S) \leq 2$};
                \node[
                    decision,
                    above=of check
                ] (check2) {$|C| = 42$};
                \node[
                    crypto,
                    left=of check2,
                    minimum width=23.5mm
                ] (decrypt) {AES-GCM$^{-1}$\\{($K$, $IV$, $C$, $A$)}};
                \node[
                    block,
                    below=5.5mm of decrypt
                ] (updatestate) {$S[A] \leftarrow (R, T^S)$};
                \node[
                    reject,
                    right=of check
                ] (reject) {Reject};
                \draw[line] (payload1) -- (timestamp1);
                \draw[line] (timestamp1) -- (counter);
                \draw[line] (counter) -- (encrypt);
                \draw[line] (encrypt) -- (concat);
                \draw[line] (concat) -- (split);
                \draw[line] (split) -- (extract);
                \draw[line] (extract) -- (checkstate);
                \draw[line] (checkstate) -- node[
                    below left=-4.5mm and 0mm,
                    font=\small\bfseries
                ]{\texttt{T}} (replay);
                \draw[line] (replay.north) |- node[
                    below left=6mm and -0.3mm,
                    font=\small\bfseries
                ]{\texttt{T}} (timestamp2.east);
                \draw[line] (replay.east) -- node[
                    below left=11mm and -0.3mm,
                    font=\small\bfseries
                ]{\texttt{F}} (reject.south);
                \draw[line] (checkstate) -- node[
                    below left=-1.9mm and -10.5mm,
                    font=\small
                ]{$A \notin S$} node[
                    below left=-1.8mm and 0.5mm,
                    font=\small\bfseries
                ]{\texttt{F}} (timestamp2);
                \draw[line] (timestamp2) |- (check);
                \draw[line] (check) -- node[
                    below left=-1.5mm and 0mm,
                    font=\small\bfseries
                ]{\texttt{T}} (check2);
                \draw[line] (check.east) -- node[
                    below left=-4mm and 0.5mm,
                    font=\small\bfseries
                ]{\texttt{F}} (reject.west);
                \draw[line] (check2.east) -| node[
                    below left=-4mm and 11mm,
                    font=\small\bfseries
                ]{\texttt{F}}(reject.north);
                \draw[line] (check2) -- node[
                    below left=-4mm and -5mm,
                    font=\small\bfseries
                ]{\texttt{T}} (decrypt);
                \draw[line] (decrypt) -- (updatestate);
                \draw[line] (decrypt) -- (payload1);
            \end{scope}
        \end{tikzpicture}}%
        \caption{
            An overview of the AEAD workflow demonstrating encryption and
            decryption lifecycle with replay protection.
        }
        \label{fig:cryptoscheme}
\end{figure}

\section{Threat model}

Our threat model assumes that traffic can be observed, modified and repeated by
a 3\textsuperscript{rd} party that is, however, unable to physically tamper with
the hardware. This reflects realistic scenario where physical access to a
satellite is rarely possible. Key rotation is out of scope and we suppose that
assets using our scheme have an appropriate electromagnetic shielding.

We presume that each asset uses a different 256-bit secret key and that IV
uniqueness is guaranteed per key by $(R \parallel T^S)$ with $R$ incremented per
frame and stored in a durable, non-volatile memory.

\subsection{Threat analysis}

In our scheme, IV serves as the reliable counter and while its predictability
might raise theoretical concerns, the monotonic nature of the timestamp and
backup counter eliminates the nonce reuse risks.

We believe that, as long as receivers retain the last seen counter and timestamp
for each AAD, a complete replay protection would be achieved in a real-world
scenario. Since the counter $R$ is incremented before each transmission, it
provides strict monotonic ordering that prevents cross-session and intra-session
replay attempts, even if an attacker can manipulate or predict timestamps. This
ensures that each message with a given counter value is accepted at most once
per asset. The defensive posture can be improved even further if receivers are
able to store and exchange this information with each other.

Provided that the IV is unique and the shared key is not disclosed, forging a
valid pair of ciphertext and tag for \mbox{AES-GCM} is computationally
unrealistic. The authentication tag covers the entire plaintext and AAD (which
is the asset ID in our case), so any modification to the message will be
detected with overwhelming probability. Even if RTC fails, IV uniqueness is
still maintained by the $R$. The time window check may cause availability
issues, but not a security failure.

Further, in regards to the authentication tag security, birthday collisions
would require approximately $2^{64}$ messages to achieve a $50\%$ probability of
finding matching 128-bit tags. At 1 Hz messaging rate with counter increments,
this would take billions of years, far exceeding any realistic mission lifetime.

The sequential nature of the IV also prevents an attacker from controlling or
predicting tag collisions, making birthday attacks computationally infeasible
within our operational constraints. An attacker attempting to XOR two captured
frames with identical IVs would require both messages to share the same counter
and timestamp values -- a remote chance scenario due to our design.

Even with intimate knowledge of mission-specific parameters, crafting a valid
forgery would require a precise timing and counter predictions, making such
attacks computationally and operationally infeasible.

\subsection{Attack scenarios}

To complement the threat model, we outline several adversarial scenarios and
describe how the proposed scheme responds to each:

\begin{enumerate}[itemsep=0pt]
    \item GNSS spoofing or time rollback -- If the on-board RTC is manipulated
        or otherwise producing incorrect timestamps, the 2 second acceptance
        window may lead to valid frames being rejected. However, the IV
        uniqueness remains guaranteed thanks to  the monotonic counter $R$.
        Additionally, ground control can temporarily perform replay checks
        relying solely on the counter until the time synchronisation is
        restored.

    \item Induced power loss -- A sudden power loss during the counter update
        could cause a state corruption. To mitigate this, for production
        implementation we strongly recommend using journaling with versioning
        and CRC. On reboot, the system would need to select the highest valid
        version and increment $R$ before transmission, thus ensuring that no IV
        reuse occurred.

    \item Replay injection -- An attacker who replays a captured frame before
        the genuine one can only do it once in each 2 second window. This does
        not imply that they can perform a forgery, therefore making replays
        significantly less valuable to a potential adversary.

    \item Cross-site replays -- If ground stations do not share the $S[A]$
        state, and DTN is in use, therefore allowing much longer acceptance
        windows, an attacker could aim to replay a valid frame at different
        stations. This risk is reduced by periodic synchronization of the $S$
        across ground sites.

    We would like to point out that even an eventual consistency significantly
        narrows the replay window and that this threat is largely immaterial in
        the regular mode of operation where ground stations verify that $(T^R -
        T^S) \leq 2$.
\end{enumerate}

We hope that these 4 scenarios illustrate that while the scheme is not designed
to prevent physical-layer attacks such as jamming, it does provide strong
guarantees against confidentiality breaches, message tampering, and replay
within the realistic operational constraints. Operational considerations are
addressed separately in the section \ref{sec:human}.

\subsection{Human factors}\label{sec:human}

While the underlying cryptographic logic is essential, we believe that
operational usability is equally important, especially in emergency situations
when human operators are required to make time-sensitive decisions based on the
limited telemetry data. Therefore, we took into account human factors to
minimise cognitive load and reduce the likelihood of operator error when
implementing or interfacing with the proposed scheme.

\begin{enumerate}[itemsep=0pt]

    \item Deterministic behaviour -- The fixed message size and single cipher
        with no optional parameters to tweak. This eliminates ambiguity,
        simplifies implementation, verification and integration.

    \item Minimal footprint -- Does not rely on CRLs or OCSP. We suggest that
        operators should not be required to debug complex workflows (such as
        PKIs) during emergencies.

    \item Clear failure modes -- Verification outcomes are non-ambiguous and
        easily interpretable:
        \begin{enumerate}[itemsep=0pt]
            \item Success -- message successfully decrypted.
            \item Failure -- decryption failed / corrupt message.
            \item Replay -- detected via $S$ or acceptance window.
        \end{enumerate}

    We aim to reduce the cognitive load and the decision ambiguity by lowering
        the probability of `partial trust' states while allowing to relax the
        acceptance window for RTC failure or DTN store-and-forward.

    Importantly, we never revert back to the plaintext transmission. This
        preserves integrity guarantees while enabling the scheme to reliably
        work across multiple scenarios.

    \item Predictable latency -- AES-GCM supported by the hardware acceleration
        allows constant-time operation, ensuring that authentication checks do
        not introduce metrics variability that can equally confuse operators and
        automated alerting systems alike.
\end{enumerate}

\noindent By prioritising simplicity, determinism and clear feedback, the
proposed scheme reduces the risk of misconfiguration or delayed response during
the mission lifetime.

\section{Implementation}

We provide a working reference implementation in the \mbox{Appendix
\ref{app:code}} (Fig.\,\ref{fig:implementation}) using Python to demonstrate the
scheme viability. Outline of the program flow is depicted in the
(Fig.\,\ref{fig:cryptoscheme}). Test vector is supplied in the (Table
\ref{tab:testvector}), Appendix \ref{app:testvector} with the complementary code
for automatic verification in the Appendix \ref{app:vectorverify}. While a
production system would likely use a different programming language and
constant-time FIPS-approved implementation that satisfies NASA-STD-1006A, our
approach lays the groundwork by validating the cryptographic design and message
structure. It makes several assumptions that align with typical spacecraft
capabilities and ground segment infrastructure:

\begin{enumerate}[itemsep=0pt]
    \item 256-bit secret key securely enrolled on the ground.
    \item AAD that is unique per deployment and asset.
    \item RTC synchronisation accuracy $\leq2$ seconds.
    \item Downlink employing FEC at the link layer.
\end{enumerate}

\section{Discussion}

Our approach prioritises simplicity and standards compliance over novel
cryptographic constructions and esoteric schemes designed to protect against
imaginary threat actors with fully functional quantum computers. By using
AES-GCM without modifications, we ensure:

\begin{enumerate}[itemsep=0pt]
    \item Visible attack surface thanks to the proven algorithms.
    \item Compatibility with existing libraries and accelerators.
    \item Reduced verification burden and overall complexity.
    \item Alignment with CCSDS recommendations \cite{ccsds3509g2}.
\end{enumerate}

This conservative approach is particularly suitable for space systems where
post-deployment updates are costly, risky and sometimes simply impossible. Our
scheme makes several deliberate compromises to balance security properties with
operational constraints:

\begin{enumerate}[itemsep=0pt]
    \item Data overhead -- the $\approx79\%$ increase in size is acceptable, as
        demonstrated in the section \ref{sec:overhead} where we quantify the
        consequences of introducing the encrypted payload compared to the plain
        SGB message.
    \item Dependency on RTC -- reasonable to assume that a satellite would have
        access to GNSS or ground link.
    \item Symmetric keys -- simpler than PKI but require secure enrolment before
        an asset is deployed on orbit.
\end{enumerate}

We propose the following areas for further exploration:

\begin{enumerate}[itemsep=0pt]
    \item Addition of the perfect forward secrecy via elliptic curve
        Diffie-Hellman ephemeral (ECDHE).
    \item Key hierarchy with separate Key Encryption Keys (KEK) and Data
        Encryption Keys (DEK).
    \item RTC recovery scenarios during system failures.
    \item Graceful key revocation and rotation procedures.
\end{enumerate}

\noindent Note should be taken that such enhancements would add complexity and,
in our opinion, should only be considered after thorough validation of the base
protocol in operational environment.

We have also studied the possibility of utilising \mbox{AES-GCM-SIV} (RFC 8452)
instead of the AES-GCM. While the former is resistant to nonce reuse, we have
decided against it due to the following reasons:

\begin{enumerate}[itemsep=0pt]
    \item Requires 2 data passes vs 1 for AES-GCM.
    \item Typically not accelerated via ARM crypto extensions.
    \item Incompatible with FIPS-140 and NASA-STD-1006A.
\end{enumerate}

\section{Conclusion}

We have presented a lightweight authenticated encryption scheme for emergency
space telemetry that balances security requirements with operational
constraints. The fixed message size ensures predictable bandwidth and simplifies
planning of link and power budgets. Hardware acceleration support for AES-GCM
available in commercial components minimises computational overhead making the
scheme suitable for deployment on satellites with comparatively modest
resources.

Our approach demonstrates that robust security does not require arcane
incantations and complex cryptographic constructions. By adhering to established
standards and focusing on implementation simplicity, we provide a scheme that
can be readily understood, implemented and validated.

We hope that our work contributes to the body of practical security solutions
for space systems, addressing the growing need for protected communications in
the domain whose capabilities sustain operations on land, sea, and in the air.

\section*{Acknowledgements}

This work expands on the materials previously created by the author for Southern
Hemisphere Space Studies Program (SHSSP) which the author was able to attend
thanks to the support from Defence SA, SASIC and ISU. The author would like to
thank Michael Bennett \mbox{(Two Swords)} along with the two anonymous experts
who conducted a review of the cryptographic design.

Additionally, the author would like to thank University of South Australia and
University of Technology Sydney library services for the access to academic
resources as well as Laurens `lvh' Van Houtven (\mbox{Latacora}) for releasing
`Crypto 101' to general public.

\bibliography{biblio}

\onecolumn

\appendix

\setcounter{section}{0}
\refstepcounter{section}
\section*{Appendix A}\label{app:code}
\,
\lstset{
    language=Python,
    basicstyle=\normalsize\ttfamily,
    breaklines=false,
    breakatwhitespace=true,
    showstringspaces=false,
    frame=none,
    numbers=left,
    numberstyle=\ttfamily\normalsize\color{grey},
    numbersep=-25pt,
    xleftmargin=-15pt,
    xrightmargin=0pt,
    aboveskip=-9pt,
    belowskip=0pt,
    commentstyle=\color{gray},
}

\begin{figure*}[htb]
    \begin{lstlisting}
        from cryptography.hazmat.primitives.ciphers.aead import AESGCM
        from secrets import token_bytes, randbits
        from time import time
        from sys import exit

        # Shared knowledge
        skey = token_bytes(32)
        cipher = AESGCM(skey)

        # Transmitter
        tx_aad = int(1234).to_bytes(2)          # 16-bit ID as AAD
        tx_ts = int(time()).to_bytes(8)         # 64-bit timestamp
        tx_ctr = int(42).to_bytes(4)            # 32-bit counter
        tx_pt = randbits(202).to_bytes(26)      # 208-bit plaintext
        tx_iv = tx_ctr + tx_ts                  # 96-bit IV
        tx_ct = cipher.encrypt(tx_iv, tx_pt, tx_aad)
        tx_frame = tx_aad + tx_iv + tx_ct

        # Receiver
        rx_state = {}
        rx_aad = tx_frame[:2]                   # 16-bit ID as AAD
        rx_ct = tx_frame[14:]                   # 336-bit ciphertext
        rx_iv = tx_frame[2:14]                  # 96-bit IV
        rx_ctr = int.from_bytes(rx_iv[:4])      # 32-bit counter
        rx_ts = int.from_bytes(rx_iv[4:])       # 64-bit timestamp

        if rx_aad in rx_state:
            last_ctr, last_ts = rx_state[rx_aad]
            if rx_ctr <= last_ctr or rx_ts <= last_ts:
                exit("REPLAY")

        if len(rx_ct) == 42 and abs(int(time()) - rx_ts) <= 2:
            rx_pt = AESGCM(skey).decrypt(rx_iv, rx_ct, rx_aad)
            print("OK") if (tx_pt == rx_pt) else exit("ERROR")
            rx_state[rx_aad] = (rx_ctr, rx_ts)
    \end{lstlisting}
    \caption{
        Intentionally simplified AEAD routine demonstrating the use of 96-bit IV
        (counter and timestamp) with 16-bit AAD. Replay tracking is not
        persisted. A production implementation would require a durable storage for
        both counter and replay state, as discussed in the paper. Tested using
        Python v3.13.5 and \texttt{cryptography} library v43.0.0 on Debian Linux
        v13.1
    }
    \label{fig:implementation}
\end{figure*}

\newpage

\refstepcounter{section}
\section*{Appendix B}\label{app:testvector}
\begin{table}[H]
    \centering
    \begin{tabular}{ll}
        \toprule
        \textbf{Parameter} & \textbf{Value} \\
        \midrule
        Key ($K$) & \texttt{1c195d64578ad0af88addd2fa452f37ee1d390728cf0258e316f1b732d2f5756} \\
        AAD ($A$) & \texttt{e802} \\
        Counter ($R$) & \texttt{7e081a3d} \\
        Timestamp ($T$) & \texttt{0eb894a953803d93} \\
        IV $(R \parallel T)$ & \texttt{7e081a3d0eb894a953803d93} \\
        Plaintext ($P$) & \texttt{e9c534097001dd986abc34454aad50bb48376c3c0de7fe3fa5ab} \\
        Ciphertext ($C$) & \texttt{62ab5d2df4687b43755b53792f9f6c6ee27169e8f89b52128cb3} \\
        Authentication Tag & \texttt{27d94586306bec73c04157efb2640c63} \\
        \bottomrule
    \end{tabular}
    \caption{
        Reference test vector for the proposed cryptographic scheme and message
        format, hexadecimal representation.
    }
    \label{tab:testvector}
\end{table}

\noindent
Verification routine:
\begin{enumerate}[itemsep=0pt]
    \item Perform AES-GCM with the key $K$, IV $(R \parallel T)$, AAD $A$, and
        plaintext $P$.
    \item Confirm that the resulting ciphertext $C$ and 128-bit tag match
        Table~\ref{tab:testvector}.
    \item Decrypt $(C \parallel \text{tag})$ using the $(K, IV, A)$ and confirm
        that the plaintext is $P$.
\end{enumerate}

\newpage

\refstepcounter{section}
\section*{Appendix C}\label{app:vectorverify}
\,
\begin{figure*}[htb]
    \begin{lstlisting}
        from cryptography.hazmat.primitives.ciphers.aead import AESGCM
        from binascii import hexlify, unhexlify

        # Vector
        K = unhexlify("1c195d64578ad0af88addd2fa452f37e" +
                      "e1d390728cf0258e316f1b732d2f5756")
        A = unhexlify("e802")                   # 16-bit ID as AAD
        R = unhexlify("7e081a3d")               # 32-bit counter
        T = unhexlify("0eb894a953803d93")       # 64-bit timestamp
        P = unhexlify("e9c534097001dd986abc34454aad50bb48376c3c0de7fe3fa5ab")
        IV = R + T                              # 96-bit IV

        # Expected outputs
        exp_tag = unhexlify("27d94586306bec73c04157efb2640c63")
        exp_frame = unhexlify(
            "e802" +                            # A
            "7e081a3d0eb894a953803d93" +        # IV = R||T
            "62ab5d2df4687b43755b53792f" +
            "9f6c6ee27169e8f89b52128cb3" +      # C
            "27d94586306bec73c04157efb2640c63"  # TAG
        )
        exp_c = unhexlify("62ab5d2df4687b43755b53792f9f6c6ee27169e8f89b52128cb3")

        # Encrypt / decrypt
        cipher = AESGCM(K)
        ct_tag = cipher.encrypt(IV, P, A)
        C, TAG = ct_tag[:-16], ct_tag[-16:]
        FRAME = A + IV + C + TAG
        pt = AESGCM(K).decrypt(IV, C + TAG, A)

        # Verify
        assert C == exp_c, f"C mismatch"
        assert TAG == exp_tag, f"Tag mismatch"
        assert FRAME == exp_frame, f"Frame mismatch"
        assert pt == P, "P mismatch"

        print("OK")
    \end{lstlisting}
    \caption{
        Test vector verification routine that validates implementation against
        the reference values from the Table \ref{tab:testvector}.\\ Tested using
        Python v3.13.5 and \texttt{cryptography} library v43.0.0 on Debian Linux
        v13.1
    }
    \label{fig:vectorverify}
\end{figure*}

\end{document}